# A Hands-On Workshop for Constructing a Low-Field MRI System in Three Days

Ivan Etoku Oiye[1], Ajay Sharma[1], Zinia Mohanta[2,3], Dinil Sasi Sankaralayam[2], Yuto Uchida[2], Teni Akinwale[4], Kexin Wang[4], Zechen Xu[4], Yifan Shuai[4], Vu Dinh[2], Sun Yuanqi[4], Aruna Singh[3], Dillip K. Senapati[2], Luke Ikard[6], Sandeep K. Ganji[7], Joseph Reilly[5], Michael Mcmahon[3], Hanzhang Lu[2], Peter Barker[4], Jennifer Morrison[5], Steven M. Ross[5], Zaver Bhujwalla[2], and Sairam Geethanath[1]*

[1]*Laboratory for Accessible MRI, Division of Cancer Imaging Research, Radiology and Radiological Sciences, Johns Hopkins University School of Medicine, 600 N. Wolfe Street, Baltimore, 21287, MD, United States*

[2]*Russell H. Morgan Department of Radiology and Radiological Science, Johns Hopkins School of Medicine, Baltimore, MD, United States*

[3]*F.M. Kirby Research Center for Functional Brain Imaging, Kennedy Krieger Institute, Baltimore, MD, United States*

[4]*Biomedical Engineering, Johns Hopkins School of Medicine, Baltimore, MD, United States*

[5]*Center for Research and Reform in Education, Johns Hopkins School of Education, Baltimore, MD, United States*

[6]*Whiting School of Engineering, Baltimore, Johns Hopkins, MD, United States*

[7]*Philips Healthcare, Baltimore, MD, USA*

**Corresponding author:**

Sairam Geethanath

sairam.geethanath@jhu.edu MRI building, 600 N. Wolfe Street,

Baltimore, MD, 21287

# Abstract

Access to Magnetic Resonance Imaging system assembly knowledge can be expanded by leveraging open-source hardware and software, simplified installation requirements, and collaborative training initiatives. To this end, we conducted a three-day workshop to construct an operational 0.27T MRI scanner. The workshop hosted 16 participants, including faculty, postdoctoral fellows, trainers, and students, who collaborated to build the scanner using open-source hardware and software components. Teams were designated to focus on various subsystems, including the magnet, passive shimming, radiofrequency (RF) coils, gradient coils, data acquisition, and reconstruction. Pre-workshop preparation involved simulation-based design processes and fabrication techniques, which incorporated configuring MaRCoS and PyPulseq libraries, CNC machining, and 3D printing. During the workshop, participants assembled an H-shaped magnet, which achieved a peak magnetic field strength of 0.269T. Passive shimming effectively reduced the field inhomogeneity from 3mT to 2mT. A 3 cm diameter RF solenoid was built and tuned to 11.4 MHz. The gradients exhibited less than 5% non-linearity in simulations and were fabricated by CNC machining copper plates. The assembled system was used to acquire a 2D spin echo of a water phantom. Following the workshop, the system was further optimized to scan relaxometry phantoms. A post-workshop survey was carried out, revealing over 87% satisfaction. The constructed scanner represents a valuable platform for educational initiatives, pulse sequence development, and preclinical research imaging efforts.

# 1. Introduction

High field MRI (1.5T - 3T) has gained wide acceptance primarily due to its superior SNR and resolution[1]. On the other hand, low-field MRI (<0.5T), presents several advantages, including open design, lower initial purchase price, operational cost, fringe fields, Specific Absorption Rate(SAR)[2], and can be tailored for specific anatomical imaging. As of 2019, two-thirds of the global population lacks access to an MRI[3]. The recent resurgence in low-field MRI[4,5] is a promising direction to address this gap. In response, academic and research groups have organized hackathons and workshops to accelerate the dissemination of low-field MRI technology, foster interdisciplinary collaboration, and cultivate open-source development communities[6]. Current initiatives have introduced open-source, very-low-field (<0.1T) MRI prototypes based on arrays of permanent magnets, particularly Halbach configurations[7]. These systems offer a more straightforward and cost-effective alternative to conventional MRI. However, they are limited by low SNR, increased susceptibility to electromagnetic noise, and prolonged acquisition time[3]. Additionally, the use of numerous small permanent magnets in their construction makes the assembly process challenging and time-consuming. These small magnets have a high surface area to volume ratio, making them prone to temperature-induced fluctuations that can degrade image quality and impair system stability[8]. An alternative magnet design has been the use of the H and C-type magnets that have been used for clinical imaging at field strengths of 0.2T, but they weighed significantly more than the Halbach arrays. The advances in the removal of the pole piece by the inclusion of shimming magnets have made this design more practical. For example, the 0.2T magnet reported in Ref.[9] is lighter than the 0.055T in Ref.[10].

Considering these recent advances in the H-type magnet designs, our team hosted a 3-day workshop in May 2025 aimed at developing an operational low-field (0.27T) H-type permanent magnet scanner with a field of view of 3 cm. The event hosted 16 participants from Johns Hopkins University and was organized by a faculty member and three students. All materials have been made publicly available as open-source packages to encourage reuse and further development. Importantly, the final scanner design is lightweight (36 kg) and modular, supporting rapid deployment in field settings. While not intended for clinical applications, it serves as a prototyping and research tool, particularly for future preclinical imaging and experimental pulse sequence development, where frequent hardware modifications are necessary.

## 2. Methods

## 2.1 Planning and preparation

The Delta DIY MRI Workshop was a hands-on initiative designed to give early-career researchers a practical introduction to MRI physics and engineering through the collaborative construction of a low-field MRI system. Preparations commenced four months prior, with weekly planning sessions to establish objectives, technical goals, materials, and contingency plans. Figure 1 presents an overview of the workshop, in which participants were organized into four teams: Magnet, RF, gradients and shims, and data acquisition and reconstruction. Each team was assigned milestone-driven tasks encompassing design, fabrication, and system integration. To ensure a smooth start, a pre-workshop "drop-in anytime" session was conducted to assist participants with the installation and configuration of Python-based tools such as PyPulseq and MaRCoS, resolving software-related issues in advance.

| Pre-workshop Planning | | Day 1 | Day 2 | Day 3 | Post-workshop |
|---|---|---|---|---|---|
| **Prelude**<br>1. Funding<br>2. Four specialized teams | Magnet | 1. Magnet simulation<br>2. Magnet assembly | 1. Field mapping<br>2. Magnet shimming | 1. Mechanical housing | **Gradient calibration**<br><br>**Phantom scanning**<br>1. Relaxometry phantom scans |
| | RF | 1. RF coil design<br>2. RF coil integration | | | |
| **Weekly meetings**<br>1. Designs<br>2. Procurements | Gradient/Shims | | 1. Passive shims design<br>2. Passive shim fabrication<br>3. Gradient design | 1. Gradient fabrication<br>2. Gradient integration | **Participants' feedback**<br>1. Testimonials<br>2. Feedback form |
| **Public outreach**<br>1. Website<br>2. Social media<br>3. Emails | Data acquisition and reconstruction | 1. Pulse Sequence Programming<br>2. MRI4all Console<br>3. Spin Echo | | 1. 2D spin echo | |
| | Milestones | 1. Spin echo | 1. Passive shimming<br>2. Gradient coil design | 1. Gradient integration<br>2. 2D spin echo<br>3. Mechanical housing | **Dissemination of results** |

*Figure 1*. Workshop overview. Pre-workshop planning involved securing funds, forming specialized workshop groups, discussing system designs, procurements, and public outreach; Day 1 of the workshop focused on magnet assembly and obtaining a spin echo; Day 2 focused on magnet characterization, passive shimming, and gradient coil design; Day 3 focused on gradient fabrication and integration, 2D spin echo acquisition, and mechanical housing; post-workshop focused on system calibration, phantom scans, Participant feedback, and dissemination of results.

Each workshop day featured a keynote on three early MRI milestones (shown in Figure 2): "The First Spin Echo," "The First MR Image," and "The First MR Scanner and Experiment," contextualizing participants' work within the field's history.

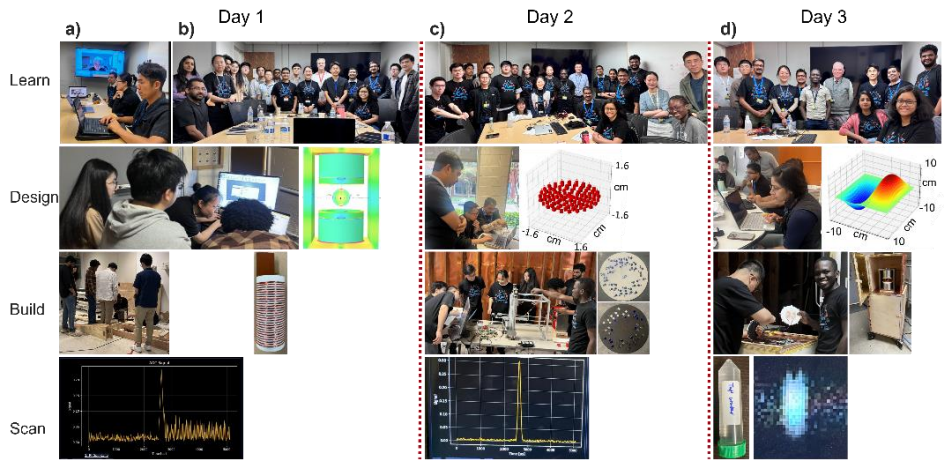

*Figure 2. Keynotes: a) Prof. Zaver Bhujwalla giving opening remarks; b) Prof. Michael McMahon, c) Prof. Hanzhang Lu, and d) Prof. Peter Barker, following their keynote lectures.*

## 2.2 The MR system:

### 2.2.1 Magnet

***Design and simulation***: Figure 3 illustrates the design and simulation of a symmetric H-magnet scanner using Opera 3D simulation software (Dassault Systemes, Inc. USA)[11]. The system comprises six cylindrical N45 NdFeB permanent magnet discs, each with a thickness of 2.54 cm and a diameter of 15.24 cm. Three magnets were assembled to form the upper section, while the remaining three formed the lower section. The yokes serve as a framework that contains and directs the magnetic flux, enhancing the efficiency of the magnetic field and concentrating the flux density in the air gap. Pole pieces were attached to the NdFeB faces to shape the magnetic field for improved homogeneity. The step-by-step design process can be accessed from GitHub[12].

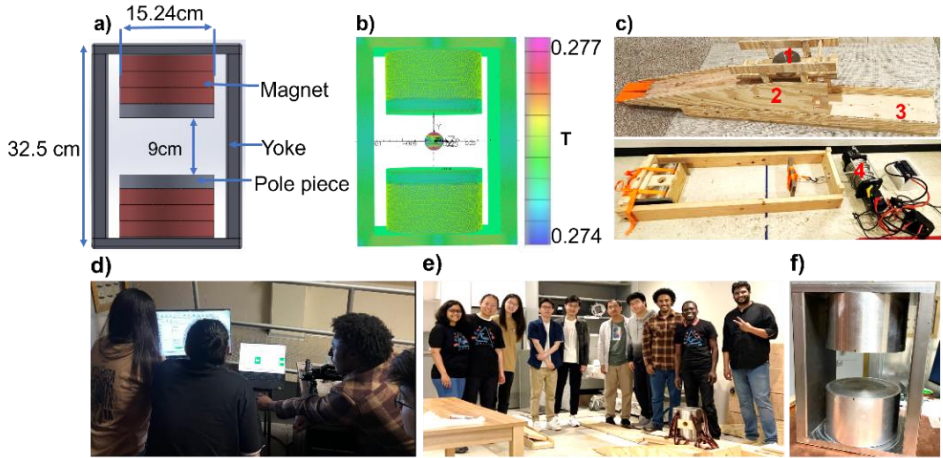

***Figure 3***. *Magnet sub-system: a) CAD model of the scanner with dimensions; b) magnetic field simulation with Opera Simulia software; c) magnet(1 highlighted in red) assembly using wooden ramp(2 highlighted in red), wooden rail(3 highlighted in red), and electronic winch(4 highlighted in red); d) participants designing and simulating the magnet; e) participants after assembling the magnet; f) assembled magnet.*

**Fabrication and integration**: We outsourced the steel component fabrication from Xometry, Inc. USA, and assembled the system using a custom wooden jig and electronic winch (Figure 3c). The step-by-step process is documented on GitHub[12]. A 9 cm space separation was secured between the pole pieces to fit the planar gradient coils, RF coil, and a 3 cm imaging field of view (FOV) was achieved. Both the yoke and pole pieces were fabricated from AISI 1018 steel.

**Field mapping:** Figure 4 shows the magnetic field characterization setup. Following the magnet assembly, we employed a home-built field mapping robot equipped with a robust three-axis linear stage, featuring a working volume of 40 x 40 x 35 cm, to measure the magnetic field [13,14]. The AlphaLab GM2 (Alphalab, Inc. USA) magnetic field probe, known for its high stability and resolution of 5 µT [15], is connected via an audio jack pin to facilitate automated measurements. All materials and source code for this field mapping robot can be found on GitHub[16].

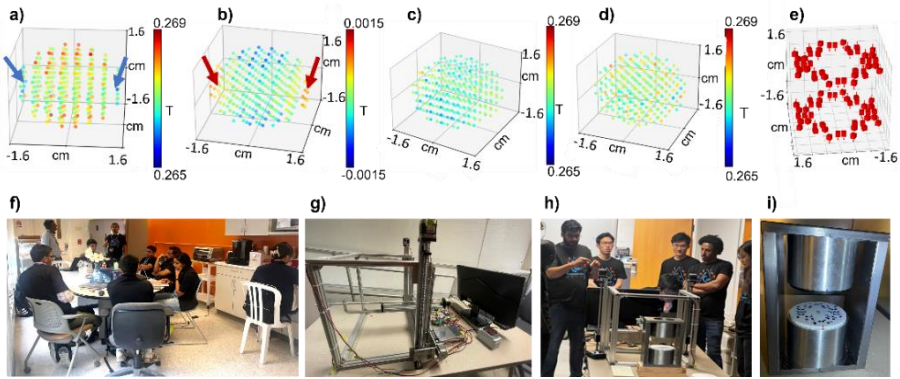

*Figure 4. Passive shimming design and integration: a) The B field measured using the field mapping robot at 0.4 cm spatial resolution for a target diameter of spherical volume of 3.2 cm, the blue arrows indicate regions of low magnetic fields; b) simulation of the resulting B field from the shim trays shows a complementary profile to the field in a), the red arrows show regions of high magnetic fields; c) the total field predicted of measured fields from the robot and the shims; d) the measured shimmed field is similar to the predicted field with increased homogeneity and reduced mean field strength (numbers in Figure 2 a); e) the output of the genetic algorithm indicates the position and polarity of the magnets to be inserted; f) participants simulating the passive shimming fields; g) is the field mapping robot; h) participants mapping the B fields of the assembled magnet with the field mapping robot; i) the 3D printed trays with magnets inserted in slots and then integrated into the magnet assembly.*

**Passive shimming:** The simulations to determine the shim magnet position and orientation (Figure 4e) were performed by implementing a genetic algorithm[17], which optimized the position and orientation of shim magnet pieces (0.6 × 0.6 × 0.6 cm) to homogenize the measured magnetic field. The optimization minimizes the ratio of the difference in magnetic field to its mean value. The simulations for the resulting magnetic field due to the shim magnets were performed using the 'magpylib' library[18]. Through the leaderboard, participants contributed to reducing the magnetic field inhomogeneity.

***Housing:*** The mechanical housing (shown in Figure 2d, row 3) was constructed from wood, while the magnet compartment was shielded using a 0.01 cm thick combination of brass and copper sheets. Additionally, 4 caster wheels (Wieyunn, Inc. China) of 12.7 cm diameter were integrated to enhance mobility[12].

## 2.2.2 RF coil and RF shield design

***RF coil:*** Figure 5 depicts the RF coil design and assembly. The orientation of the $B_0$ field in an H-shaped magnet design allows the use of a solenoid geometry for the RF coil[19]. We designed and built a transmit-receive solenoid coil with an inner diameter of 3 cm, an outer diameter of 3.4 cm, and a length of 6.8 cm. It features a 0.25 cm pitch and is wound with copper wire (AWG 16). The coil is segmented into two by an 8pF ceramic capacitor (Murata Electronics, Inc. Japan) to distribute the electric field, and phase shifts along the length of the conductor[20]. It was tuned to an operating frequency of 11.4 MHz and impedance matched to 50 Ω using two variable capacitors (Knowles Voltronics, Inc. USA)[12].

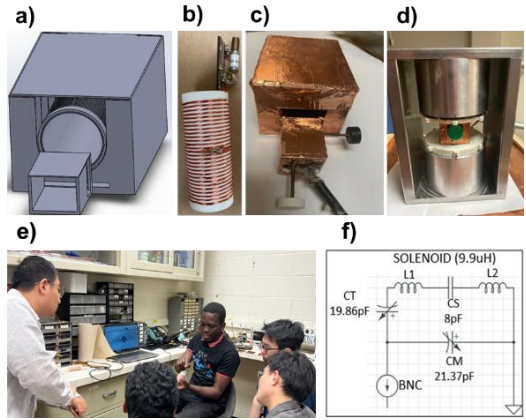

*Figure 5. RF sub-system: a) CAD model of the RF coil in an RF shield; b)RF solenoid; c) constructed RF in the RF shield; d) RF and RF shield integrated to the scanner assembly; e) participants designing RF and the RF shield; f) RF circuit diagram showing the tuning capacitor (CT), matching capacitor (CM), solenoid segmented into two (L1 and L2) by ceramic capacitor(CS).*

**RF shield:** A 0.01cm layer of solid copper tape was applied to a 5.6 x 6.8 x 9.8 cm 3D printed case to reduce electromagnetic interference (EMI) by minimizing the interaction between the RF coil and its external environment (illustrated in Figure 5c). Positioning the RF coil in close proximity to the shield introduced negative mutual inductance into the RF coil circuit, which in turn raised the transmit frequency. Therefore, it was crucial to incorporate the RF shield during coil frequency tuning and impedance matching[6].

## 2.2.3 Gradients:

***Design and simulation:*** Figure 6 illustrates the design process for the magnetic gradient coils. The desired magnetic gradient field parameters, such as the maximum limit of 5% non-linearity, two directions, and maximum gradient strength, were defined and instantiated with coil radius, current, and mesh resolution. Similar to the passive shimming implementation, the gradient coil design also involved the use of genetic algorithms, and based on the use of the current density function modeled and modulated by a weighting function, with the resulting contours being the wire patterns for the gradient coil, similar to ref[21]. The source code for gradient coil design, including parameter values for the genetic algorithm implementation, is available online on the GitHub repository[22]. A sample demonstration is also provided on the corresponding Wiki page[12]. Similar to the passive shimming leaderboard, the participants participated in reducing the non-linearity of the gradient design by modifying the genetic algorithm parameters.

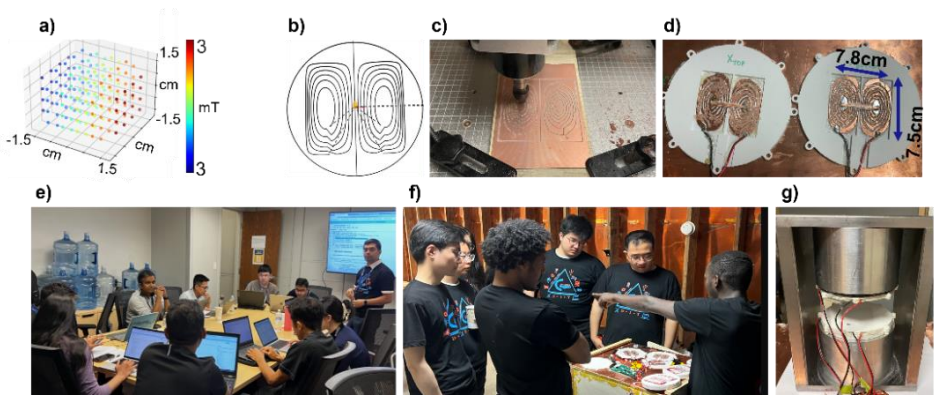

*Figure 6.* Gradient sub-system: a) Gradient field simulation using magpylib; b) contours showing patterns to be etched; c) CNC machine etching contour patterns from a 0.1cm thick copper plate; d) gradients glued to the 3D printed gradient holder; e) participants designing and simulating gradients; f) participants integrating the CNC machined gradient coils to the gradient holders; g) gradients integrated to the magnet assembly with plastic screws reinforced with glue.

**Fabrication:** The contours from the simulations (shown in Figure 6b) were converted into G-code (code available in Ref.[22]) and uploaded into a CNC machine (CNCTOPBAOS, Inc. China), which etched the patterns onto a 0.1 cm thick copper plate. The resulting pieces were then bonded to 3D-printed holders with plastic glue to prevent movement due to torque during gradient switching. Subsequently, the gradients were secured to the magnet assembly with plastic screws and connected to the GPA outputs. Due to the lengthy CNC fabrication process (7 hours for one gradient pair), we were limited to two gradient sets (X and Y).

*Gradient coil calibration:* Figure 7 illustrates the calibration setup for the gradient coils. A 3D printed jig was designed to hold the gradients in position, mimicking the gradient configuration as if it were in the scanner. We then connected each gradient pair to 10 resistors (Jersvimc, Inc. China), each of 10 watts in parallel, and supplied 6 amps, 4.72 V from a DC power supply. A field mapping robot was used to measure the gradient field for each gradient pair across 4 cm with a resolution of 0.5 cm along the center of the gradients. The process was repeated with 8 amps, 6.24 V.

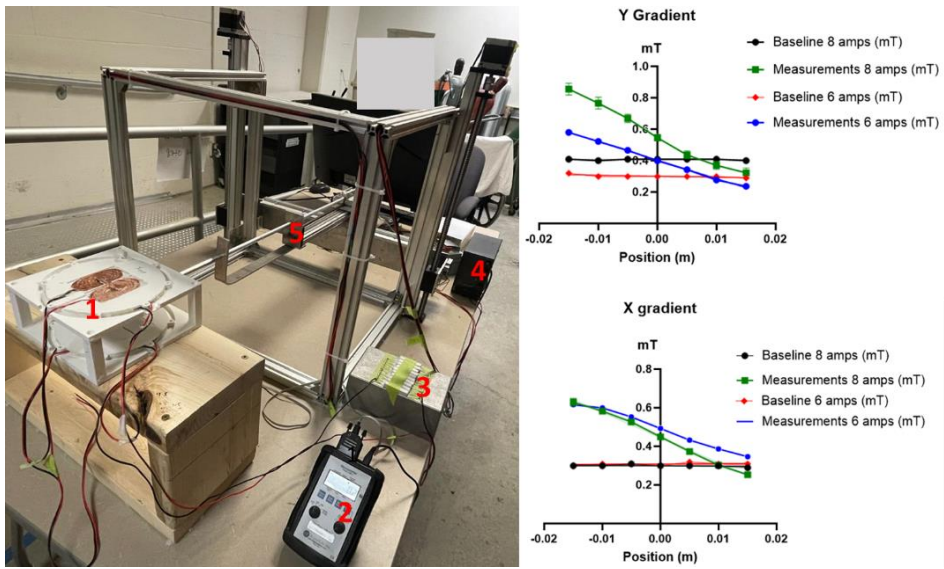

*Figure 7. Gradient characterization: Gradients held in place with a 3D printed jig (1 highlighted in red), connected to 10 resistors (3 highlighted in red) in parallel, powered by a DC power supply (4 highlighted in red) with 8 amps, 6.24V, and 6 amps, 4.72V. A field mapping robot (5 in red color) connected to a Gauss meter (2 highlighted in red) measured the*

*gradient field across 4 cm with a resolution of 0.5 cm along the center. X Gradient efficiency of 1.630625 mT/M/A, X Gradient max of 16.30625 mT/M/A, Y Gradient efficiency of 2.1525 mT/M/A, and Y Gradient max of 21.525 mT/M/A (with a 10A amplifier) were calculated from the measurements using first-order spherical harmonic fitting.*

### 2.2.4 Amplifiers

***Gradient amplifier:*** We utilized an open-source gradient amplifier developed by Cooley et al.,[23], which operates on ±15V supply rails. This amplifier utilizes OPA549 (Texas Instruments, Inc. USA) operational amplifiers arranged in a push-pull configuration to drive the gradient coils efficiently. It can deliver a minimum of 5 A of current with an approximate bandwidth of 10 kHz. A feedback loop, incorporating a current-sensing resistor, ensures a stable output by maintaining the current in proportion to the control voltage input. This configuration supports reliable and consistent gradient switching, making it ideal for both imaging applications and educational purposes.

***TX amplifier:*** We utilized the Mini-Circuits ZHL-3A 29.5 dBm (1 W) amplifier, featuring 24 dB of gain, to amplify the transmit pulses. Details of this can be accessed from the ref. [23]

***RX amplifier:*** We utilized a 2-stage low-cost (< $3 USD) GALI-74+ preamplifier ICs (Mini-Circuits, Inc. USA) with a 2.7 dB noise figure and 25 dB of gain to amplify the received signal[23] . A T/R switch with a classic PIN diode quarter-wave design was integrated and controlled by the console's TX gating signal, laid on a single PCB with the preamplifier.

### 2.2.5 Spectrometer

We utilized the Red Pitaya SDRlab 122-16 board as the spectrometer, controlling scanner components through DAC and ADC channels, as well as the SPI bus to interface with the GPA-FHDA gradient amplifier. The system connects to a host computer via Ethernet and supports custom firmware. It employs the open-source MaRCoS library[24], which includes both firmware and a Python interface for executing MRI sequences and acquiring data. Additionally, it integrates with the FLOCRA-Pulseq interpreter to convert Pulseq-formatted sequences into MarCos instructions, facilitating sequence design through PyPulseq. The FLOCRA-Pulseq interpreter was refactored and updated before the workshop to be compatible with the latest version of Pypulseq (1.4.2).

### 2.2.6 Console

We used the MRI4ALL console software developed during the 2023 MRI4ALL Hackathon, an open-source platform written entirely in Python 3 and utilizing PyQt5 for its graphical user interface[25]. The software runs on Ubuntu 22.04 and comprises three modular services: scan planning and visualization, scanner control, and image reconstruction. Post-workshop, we hosted the console on an Intel Next Unit of Computing (NUC) device to reduce the form factor of the overall scanner.

### 2.2.7 Data acquisition and reconstruction

We used the Pypulseq library[26] to generate pulse sequence waveforms and timings that were stored in '.seq' files that were interpreted by the Marcos spectrometer[24]. Two sequences were employed to characterize and debug the system: (1) An adjust frequency finder using an RF spin echo pulse (SE) sequence that rasters through a range of transmit frequencies to find the resonant frequency with the highest SNR. Imaging included the use of 2D SE sequence to accomplish two contrasts: $T_1$w (TR/TE: 250/3 ms) and PD (TR/TE: 5000/3 ms). Due to the inhomogeneity (~2 mT) of the magnetic field, scanning with long TE (>20 ms) introduced significant noise, which limited us from running $T_2$w scans. The characterization sequences with similar acquisition parameters have been described recently[27].

## 2.2.8 MR scanning

*Preliminary test:* During the workshop, a 3 cm test tube filled with tap water (Figure 2d, row 3) was scanned with TR/TE: 5000/3 ms, NSA: 1, BW:16,000 Hz, FOV: 64, and a base resolution of 32.

*Contrast:* Figure 8 illustrates 2 cm spherical samples containing ultrapure water and $CuSO_4$ solutions at concentrations of 6 mM, 20 mM, 40 mM, and 60 mM, each prepared in ultrapure water. $T_1$w images were acquired using a 2D SE sequence with TR/TE: 250/3 ms, NSA: 32, BW: 16,000 Hz, FOV: 64, and a base resolution of 32. Proton density (PD) images were acquired using a 2D SE sequence with TR/TE: 5000/3 ms, NSA: 1, BW:16,000 Hz, FOV: 64, and a base resolution of 32.

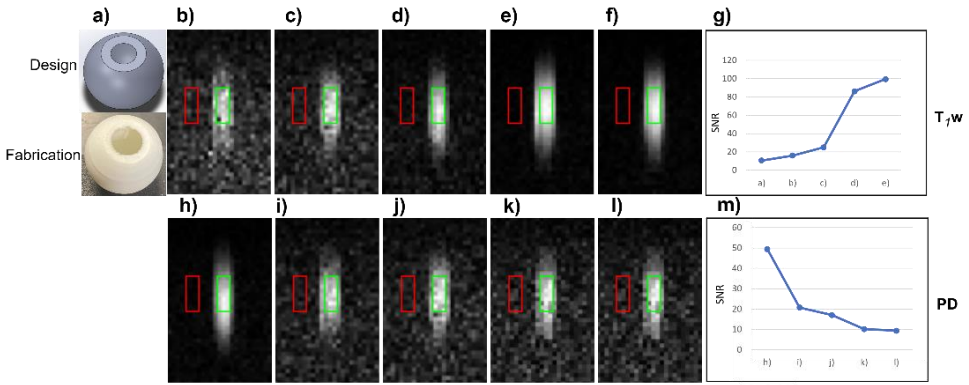

*Figure 8*. Spherical phantom scans. a) 2 cm spherical phantom; $T_1$w image of b) water, c) $CuSO_4$ 6mM, d) $CuSO_4$ 20mM, e) $CuSO_4$ 40mM, f) $CuSO_4$ 60mM scanned at TR/TE: 250/3, NSA 32, BW 16000, FOV 64, base resolution 32. The green boundaries demarcate the ROI in the object while the red boundaries demarcate the background noise, g) SNR increasing from water(b) to $CuSO_4$ 60mM(f); PD images of: h) water, i) $CuSO_4$ 6mM, j) $CuSO_4$ 20mM, k) $CuSO_4$ 40mM, l) $CuSO_4$ 60mM, scanned at TR/TE 5000/3, NSA 1, BW 16000, FOV 64, base resolution 32; m) SNR

decreasing from water(h) to CuSO₄ 60mM(l).

**Spatial resolution**: A 2 cm spherical phantom (Figure 9a) was partitioned into 50% and 30% sections, both filled with 60 mM CuSO4 and scanned with the parameters: TR/TE: 250/3 ms, NSA: 32, BW: 16000, FOV: 64, base resolution: 32. Additionally, a 2.8 cm thick cylindrical phantom (Figure 9b) with two cylindrical extrusions measuring 2.3 cm each, both filled with 60 mM CuSO4, was scanned using TR/TE: 250/3 ms, NSA: 99, BW: 16000, FOV: 64, and base resolution of 32

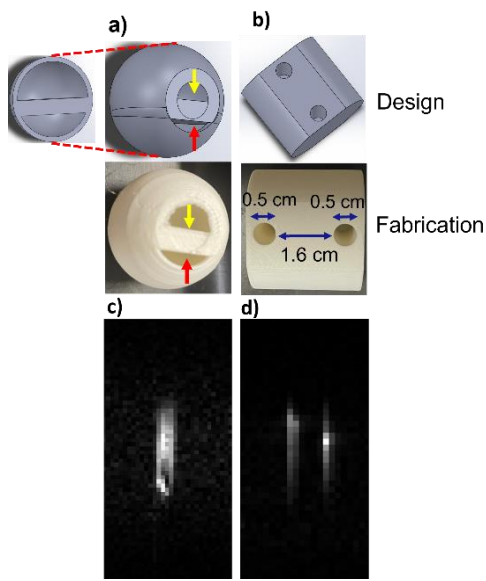

*Figure 9. Partitioned phantom scans. a) 2 cm phantom partitioned to 50% and 30% (the red arrows show 50%, and the yellow arrows 30%) filled with CuSO₄ 60mM in both partitions; b) 2.8 cm thick cylindrical phantom with 2 cylindrical extrusions of 2.3cm, both filled with CuSO₄ 60mM; c) 2D SE of CuSO4 60mM of phantom a) scanned at TR/TE: 250/3, NSA 32, BW 16000, FOV 64, base resolution 32; d) 2D SE of CuSO₄ 60mM of phantom b) scanned at TR/TE: 250/3, NSA 99, BW 16000, FOV 64, base resolution 32, showing spatial resolution.*

## 3. *Workshop survey*

A post-workshop survey was conducted to assess participant satisfaction across four primary domains: Learning & Skill Transfer, Technical Engagement, Workshop Delivery, and Social Dynamics. This comprised a set of ten questions administered using a Google Forms webpage. These questions were formulated in consultation with the collaborators from the Johns Hopkins School of Education, with a primary objective of understanding if the participants were able to accomplish building a scanner with target specifications in three days.

## 4. Results

### *MR system*

**Magnet assembly:** The magnet assembly yielded a peak magnetic field of 0.269T at the isocenter, with a 3mT field inhomogeneity before passive shimming, and 2mT field inhomogeneity after passive shimming, across a 3.2 cm DSV measured at a resolution of 0.4 cm, mapped using a 3D field mapping robot. This was close to the simulated value of 0.277T and inhomogeneity of 3mT. Manufacturing tolerances and flexing of the final structure, as well as variation in the properties of individual magnets, are the likely cause for modest differences between the simulated and actual field measurements. The measured field data from the field mapping robot was instrumental in determining the position and orientation of the shim magnets. Subsequently, we incorporated 84 small magnets (0.625 cm³) in a 3D printed tray for shimming, which successfully reduced the inhomogeneity to 2mT.

**Gradients**: The measured resistance for both the X and Y gradients was 0.4 $\Omega$, and the inductances of X and Y gradients were 35.8 µH and 38.864 µH, respectively. Table 1 shows the slope and efficiency of the x and y gradients.

**Table 1:** *slope and efficiency of the x and y gradients.*

| Strength | Gx (mT/m) | Gy (mT/m) |
|---|---|---|
| 6amps, 4.72V | -9.69 | -11.76 |
| 8amps, 6.24V | -13.17 | -18.76 |
| Efficiency | Gx (mT/m/A) | Gy (mT/m/A) |
|  | 1.63 | 2.15 |

The maximum gradient strengths (16.3mT/m/A for x-gradient and 21.5 mT/m/A for y-gradient) were calculated by first-order spherical harmonic fitting using a 10A gradient amplifier.

***RF Coil:*** The RF coil measured an inductance of 9.94 µH and revealed a reflection coefficient of –35dB at 11.4 MHz. Details of the coil can be accessed in Ref. [12]

***Phantom scans:*** The RF Spin echo (SE) and 2D SE of the 3 cm water phantom obtained during the workshop are shown in Figure 2 (row 3). The boost in SNR in the RF SE was enhanced by passive shimming. Figure 8 shows spherical phantom scans of water, $CuSO_4$ 6mM, $CuSO_4$ 20mM,

$CuSO_4$ 40mM, and $CuSO_4$ 60mM. In $T_1$w images, phantoms with higher concentrations of $CuSO_4$ solutions, appear brighter relative to the background, with SNR increasing with $CuSO_4$ concentration (left to right), while in PD images, phantoms with less $CuSO_4$ concentration appear brighter with SNR decreasing with increasing concentration of $CuSO_4$ (left to right). Figure 9 shows a $T_1$w 2D image of $CuSO_4$ in partitioned phantoms, revealing spatial resolution. The images are distorted in shape due to the inhomogeneity of the magnet and the non-linearity of the gradients.

***Participant experience and feedback***:

The results of the workshop survey, presented in Figure 10, demonstrate that the workshop met its learning objectives. In the learning & skill transfer category, every participant reported acquiring new knowledge, with 43.75% indicating they learned significantly. Furthermore, 56.25% of attendees expressed confidence in their ability to reproduce their work independently or collaboratively using the resources provided, while 43.75% responded "Maybe," indicating a strong foundation for future skill application.

In the technical engagement, 87.5% of participants reported experiencing hands-on learning, and an equivalent percentage successfully assembled parts of a scanner, underscoring the workshop's focus on practical implementation. Additionally, 75% indicated that the subsystem they worked on met its technical specifications.

The workshop delivery section yielded strong results, with 87.5% affirming that the workshop met all the committed milestones. When asked about their likelihood of attending future workshops, 62.5% expressed a definite interest in returning, while 31.5% responded "Maybe," reflecting a general satisfaction with the structure and delivery.

Lastly, the social dynamics category received exceptionally positive feedback; 93.7% of participants felt that the team environment, peer interactions, and support from organizers met their expectations, and the same proportion found the overall atmosphere to be respectful and collaborative. Equally high numbers indicated they would recommend the workshop to their peers. Collectively, these results affirm the workshop's design as a replicable and high-impact model for experiential scientific training.

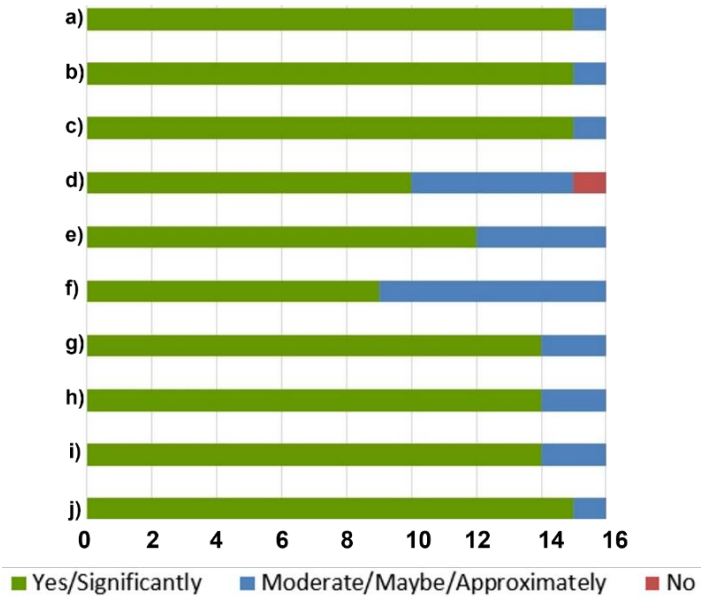

*Figure 10.* Feedback from participants on the workshop experience. a)Did you learn anything new? b) Did you experience hands-on learning? c) Did you experience putting parts of a scanner together? d) Did the workshop deliver on the committed milestones? e) Can you reproduce your work with the workshop resources? f) Did the subsystem meet specifications? g) Will you attend the same workshop again? h) Will you recommend this workshop to peers? i) Were the social aspects as expected? J) Was the ambiance respectful and collaborative?

## 5. Discussion

***Post workshop:*** After the workshop, we designed and constructed a larger trans-receive RF coil with an inner diameter of 3.4 cm, an outer diameter of 3.8 cm, and a length of 7.6 cm, featuring a pitch of 0.25 cm made from Litz wire (AWG 16). The coil was segmented into four parts using surface mount capacitors and tuned to 11.475 MHz, achieving an impedance match to 50 ohms, with a reflection coefficient of -45 dB. We incorporated a sample slide to help hold and position the sample at the isocenter. Despite the benefits of Litz wire in reducing sheath and proximity effects due to its construction from multiple thin insulated wire strands, each with a diameter smaller than the skin depth [19], the coil did not function as intended; its effectiveness has been generally demonstrated for RF coils up to 10 MHz [28]. This can also be attributed to the close proximity of the RF shield to the RF coil, reducing the transmit efficiency, with a theoretical value of zero if the shield coincides with the RF coil. This topic is extensively covered on pages 364–371 by Mispelter et al. [29]. So, we had to revert to the coil built during the workshop.

### Comparison with other workshops

*Table 2: Comparison of four MRI hackathons and workshops: P.O.C. MRI in Africa (Obungoloch et al., 2022), MRI4All Hackathon (Block et al., 2023), ezyMRI Hackathon (Hung et al., 2024), and the three-day Workshop 2025. The table shows duration, field strength, scanner type, software/console, training structure, and participant contributions. Collectively, these initiatives illustrate diverse models for capacity building, technical training, and open-source development in accessible MRI.*

| Feature | P.O.C. MRI in Africa (Obungoloch et al., 2022) | MRI4All Hackathon (Block et al., 2023) | ezyMRI Hackathon (Huang et al., 2024) | Delta Workshop 2025 |
|---|---|---|---|---|
| Duration (days) | 11 | 7 | 4 | 3 |
| Field strength (T) | 0.046T | 0.043T | 0.038T | 0.27T |
| Open-source material | N/A | GitHub | N/A | GitHub |
| Scanner type | Halbach | Halbach | Halbach | H-Magnet |
| Software / Console | 1. Red pitaya 122-16<br>2. OCRA | 1. Red pitaya 122-16<br>2. MaRCoS GUI | 1. Red pitaya 122-16 board<br>2. MaRGE GUI | 1. Red pitaya 122-16<br>2. MRI4all GUI<br>3. MaRCoS |
| Structure | 1. Online training<br>2. hands-on training | 1. Structured Teams<br>2. Weekly meetings<br>3. hands-on training | 1. Pre-workshop lectures<br>2. Structured Teams<br>3. hands-on training | 1. Structured Teams<br>2. Keynote Lectures<br>3. Hands-on training |
| Participant contribution | 1. Hardware | 1. Hardware<br>2. Software | 1. Hardware<br>2. Software | 1. Hardware<br>2. Software |

The increasing interest in open-source, low-field MRI platforms has resulted in hands-on workshops and hackathons aimed at democratizing MRI technology (shown in Table 2). Halbach arrays have been particularly popular due to their reduced stray fields and inherently safer assembly process.

However, their complex construction requiring precise placement of numerous small magnets and limited field strengths (<0.1T) constrains their utility, especially for applications requiring higher signal fidelity and spatial resolution. In comparison, our H-magnet design achieves a higher field strength of 0.27T, which translates to improved SNR and more robust anatomical imaging, even within a compact footprint. These characteristics broaden its potential for applications in small animal imaging and hardware prototyping.

However, the benefits of higher magnetic field strength come with safety challenges during construction. The assembly of large permanent magnets can be hazardous due to the significant attractive forces involved. To mitigate these risks, we developed a mechanical jig (Figure 3c) comprising a wooden ramp and an electric winch capable of delivering a pulling force up to 13,500 lb. This strategy, while effective, necessitates careful alignment of the magnets post-assembly to maintain homogeneity, typically achieved through mallet-based adjustments. Such alignment procedures introduce variability and may benefit from future automation or jacking systems to further enhance safety and precision. Unlike Halbach designs, H-magnet configurations involve a ferromagnetic yoke, which complicates simulation using common open-source software like Magpylib [18] and Magtetris [30]. As a result, we employed Opera 3D simulation software (Dassault Systemes, Inc. USA) [11], a commercial finite element simulation package to model both the magnetic field distribution. The lack of open-source alternatives for yoke-based magnet simulations remains a critical gap in the low-field MRI development ecosystem.

In terms of system control, the Red Pitaya 122-16 board remains central to many open-source MRI consoles such as OCRA and MaRCoS [31]. Although the Red Pitaya community continues to contribute actively to the development of these platforms, the discontinuation of the 122-16 model raises concerns about long-term sustainability and availability. Future work should explore alternative FPGA-based and software-defined radio solutions that preserve compatibility with existing open-source console software or facilitate transitions to new architectures with minimal re-engineering effort.

Importantly, the final scanner design is both lightweight (36 kg) and modular, supporting rapid deployment in field settings. While not intended for clinical applications, it serves as a powerful prototyping and research tool, particularly for small animal imaging and experimental pulse sequence development, where frequent hardware modifications are necessary.

## 6. Conclusion

We successfully conducted a three-day workshop that enabled participants to access knowledge to simulation, design, fabrication and assembly of subsystems in an MR scanner, in a hands-on manner. The DIY MRI scanner was able to generate in vitro phantom images reflecting the ability to visualize changes in contrast and spatial distribution. Post-workshop refinements improved the system but it still requires volume-based shimming using gradient coils or additional shim coils. The participants provided a positive response to the content, structure, and take-home resources. Future events of this workshop can also enable establishing consensus on standards for open-source low-field MRI subsystem design and fabrication.

## 7. Data availability

This article demonstrates the feasibility of constructing a low-field MRI system within a three-day workshop. All datasets referenced and summarized are publicly accessible from the online sources cited in the references. Readers may obtain the datasets and source codes directly from the repositories cited throughout the article.

## 8. Acknowledgments

This research was funded by the Johns Hopkins Provost's DELTA Award (PI: Geethanath) and Medirays Healthcare Pvt. Ltd (PI: Geethanath).

## 9. Conflict of interest statements

The authors have no conflicts of interest to disclose.

## 10. Author contribution

| | |
|---|---|
| Conceptualization | Ivan Etoku Oiye, Sairam Geethanath, Zinia Mohanta, Ajay Sharma, Dinil Sasi Sankaralayam |
| Methodology | Ivan Etoku Oiye, Luke Ikard, Zinia Mohanta, Ajay Sharma, Zechen Xu, Yuto Uchida, Teni Akinwale, Yifan Shuai, Sun Yuanqi |
| Software | Sairam Geethanath, Ajay Sharma, Vu Dinh, Aruna Singh, Sandeep K. Ganji, Dillip K. Senapati, Kexin Wang |
| Validation | Steven M. Ross, Joseph Reilly, Jennifer Morrison, Sairam Geethanath |
| Formal analysis | Ivan Etoku Oiye, Zinia Mohanta, Ajay Sharma |
| Investigation | Ivan Etoku Oiye, Zinia Mohanta, Sairam Geethanath |
| Resources | Zinia Mohanta, Sairam Geethanath |
| Data Curation | Ivan Etoku Oiye, Zinia Mohanta, Ajay Sharma |
| Writing - Original Draft | Ivan Etoku Oiye, Zinia Mohanta, Ajay Sharma, Sairam Geethanath |
| Writing - Review & Editing | Ivan Etoku Oiye, Zinia Mohanta, Ajay Sharma, Dinil Sasi Sankaralayam, Sairam Geethanath, Yuto Uchida |
| Visualization | Zaver Bhujwalla, Peter Barker, Hanzhang Lu, Michael McMahon |
| Supervision | Ivan Etoku Oiye, Zinia Mohanta, Ajay Sharma, Sairam Geethanath |
| Project administration | Sairam Geethanath |
| Funding acquisition | Sairam Geethanath |